\documentclass{osa-article}
\journal{oe}
\usepackage{graphicx}
\usepackage{mathrsfs}
\usepackage{bm}
\usepackage{amsmath}
\usepackage{dcolumn}
\usepackage{color}
\usepackage{siunitx}
 
\def\u#1{_{\rm #1}}
\newcommand{\ket}[1]{| #1 \rangle}

\newcommand{\expect}[1]{\langle #1 \rangle} 
\def\H{{\rm H}}
\def\V{{\rm V}}

\def\v{\u{V}}

\def\00{\H\V}
\def\11{\V\H}

\DeclareRobustCommand{\erase}{\bgroup\markoverwith{\textcolor{red}{\rule[.5ex]{2pt}{0.4pt}}}\ULon}

\def\piyo{black}
\def\nyan{black}

\begin{document}
\title{Ultra-fast Hong-Ou-Mandel interferometry via temporal filtering}

\author{
Yoshiaki~Tsujimoto,\authormark{1} Kentaro~Wakui,\authormark{1} 
Mikio~Fujiwara,\authormark{1}  Masahide~Sasaki,\authormark{1} Masahiro~Takeoka,\authormark{1,2} 
}

\address{\authormark{1}Advanced ICT Research Institute, National Institute of Information and Communications
Technology (NICT), Koganei, Tokyo 184-8795, Japan}
\address{\authormark{2}Department of Electronics and Electrical Engineering, 
Keio University, Yokohama, Kanagawa 223-8522, Japan}

\email{\authormark{*}tsujimoto@nict.go.jp} %% email address is required

\begin{abstract} 
Heralded single photons~(HSPs) generated by 
spontaneous parametric down-conversion~(SPDC) are useful
resource to achieve various photonic quantum information processing. 
Given a large-scale experiment which needs multiple HSPs, 
increasing the generation rate with suppressing higher-order pair creation is desirable. 
One of the promising ways is to use a pump laser with a GHz-order repetition rate. 
In such a high repetition rate regime, however, single-photon detectors can only partially identify the pulses.
Hence, we develop a simple model to consider that effect on the spectral purity, and experimentally demonstrate a high-visibility Hong-Ou-Mandel interference between two independent HSPs generated 
by SPDC with 3.2\,GHz-repetition-rate mode-locked pump pulses. 
The observed visibility of $0.88(3)$ is in good agreement with our theoretical model.
%Our method forms an important building block to achieve large-scale, high-fidelity, and high-speed photonic quantum information processing. 
\end{abstract}
%\pacs{42.65.Lm, 42.50.−p}

%\ocis{(060.5565) Quantum communications; (270.0270) Quantum optics.} 

\section{Introduction}
%{\it Introduction.-}
Photon pairs generated by spontaneous parametric down-conversion~(SPDC) 
are essential resource for photonic quantum information processing, 
such as fundamental tests of nonlocality~\cite{giustina2013bell,PhysRevLett.115.250402}, 
quantum communication over long distance~\cite{Azuma2015,Yin2017,Wengerowsky2020}, quantum metrology~\cite{Nagata2007,Slussarenko2017}, and boson sampling~\cite{10.1145/1993636.1993682,PhysRevLett.121.250505}. 
To push these technologies forward, there are two basic requirements for the SPDC sources: high quality of the photons and high \textcolor{\piyo}{clock} rate for their generation. For the quality, tremendous efforts have been devoted so far. 
In light of the recent progresses on developing SPDC sources with high brightness, indistinguishability, and collection efficiency, the 
design of the source itself is approaching to the optimal one~\cite{PhysRevLett.121.250505,Meyer-Scott:18}. 
For the \textcolor{\piyo}{clock} rate, it had been limited by the repetition rate of pump lasers \textcolor{\piyo}{mostly driven in the range of tens of MHz due to conventional laser oscillator designs.} 
Recently, \textcolor{\piyo}{GHz-repetition-rate mode-locked pump lasers were introduced to demonstrate ultra-fast generation of SPDC photon pairs at telecom wavelengths}~\cite{Zhang:08,Jin2014,Ngah2015} based on the photonics technology, by which the clock rate has reached to 50\,GHz~\cite{Wakui:s},
and very recently, the Hong-Ou-Mandel~(HOM) interference~\cite{PhysRevLett.59.2044} of the heralded single photons (HSPs) from independent SPDC sources with a GHz-order pump was also observed~\cite{d2020universal}.
%One can think of using even faster mode-locked lasers for pumping. 
%Then a technical but essential question arises: is it possible to further increase the rate of photon-pair generation by simply increasing the pump rate? Or is there any intrinsic limiting factor?

\textcolor{\nyan}{To observe the high-visibility HOM interference using such 
ultra-fast photon pair generation systems, employing narrow detection windows~(temporal filtering)~\cite{Halder2007,Tsujimoto2017,Tsujimoto2018,Miyanishi2019,Samara2020} 
is necessary, otherwise the adjacent pulses are accidentally considered as a successful event. 
Although this condition has been necessary in previous experiments, 
its influence on the interference visibility has not been argued, 
since the typical pulse interval was much larger than the timing resolution of single-photon detectors. 
On the other hand, when the pump repetition rate is high, the single-photon detectors can only partially identify the pulses.  
In such a regime, a theoretical model to deal with the influence of the pump repetition rate 
and timing resolution is required.} 

\textcolor{\nyan}{In this paper, we incorporate the influence of the pump repetition rate and timing resolution of detectors 
into the joint spectral amplitude~(JSA) of the photon pair. 
In the model, the detection of adjacent pulses emerges as a 
comb structure of the JSA, which enables us to quantitatively estimate the 
spectral purity of the photon pair from experimental parameters. 
Moreover, we demonstrate the high-visibility HOM interference 
between two independent HSPs generated by SPDC with 3.2\,GHz-repetition-rate mode-locked pump pulses 
via temporal filtering. To our knowledge, this is the highest repetition rate with an observation of the high-visibility HOM interference between the HSPs. 
The observed visibility of $0.88(3)$ is in good agreement with the theoretical value of 0.94. 
These results will be useful for a large-scale, high-fidelity, and high-speed photonic quantum information processing.}

\begin{figure}[t]
 \begin{center}
%\scalebox{0.27}{\includegraphics{./BellSetup.pdf}}
 \includegraphics[width=\columnwidth]{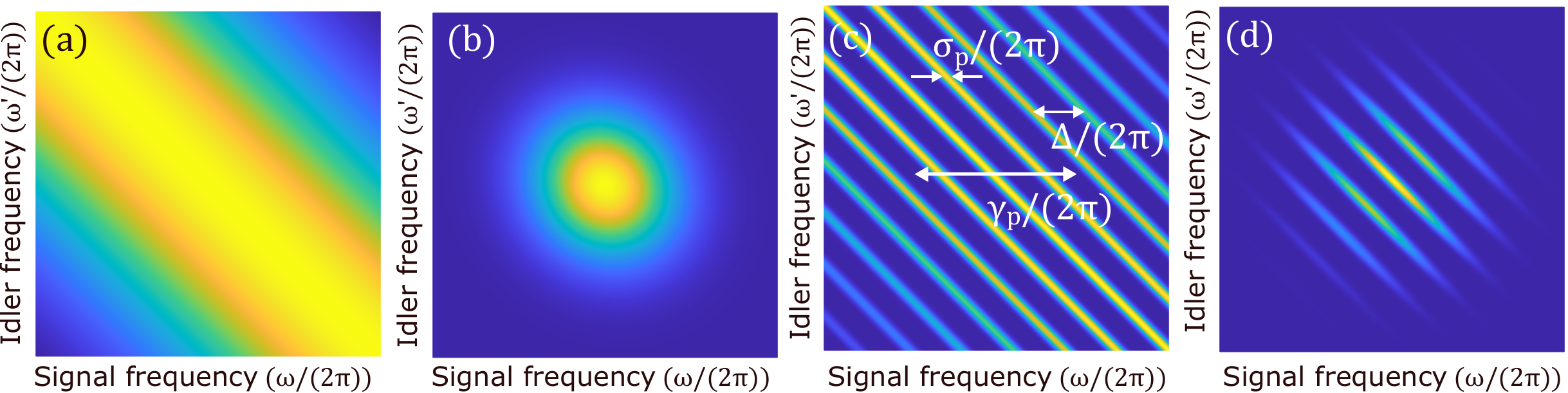}
  \caption{(a)~The JSA of the biphoton state with pulsed pump. The horizontal and vertical axes represent frequencies of the signal and idler photons, respectively. 
  (b)~The JSA after performing GVM or frequency filtering on  (a). (c)~The JSA of the biphoton state considering the comb structure of the pump. (d)~The JSA after performing GVM or frequency filtering on (c). 
  \label{fig:SPDC}} 
 \end{center}
\end{figure}

%{\it Pump repetition rate and purity of the HSP.-} 
\section{Pump repetition rate and purity of the HSP}
\label{sec1}
We first revisit the method to generate the pure HSP, and then show 
the influence of the pump repetition rate on the spectral purity. 
The spectral purity of the HSP is calculated from the JSA of the biphoton state~(the detail is given in Appendix~\ref{app:A}). In pulsed pump regime, the JSA can be assumed to be a broad two-dimensional~(2D) Gaussian, which inherits the spectral distribution of the mode-locked pump pulse as shown in Fig.~\ref{fig:SPDC}(a). 
Here, for simplicity, we \textcolor{\piyo}{assumed that the phase matching bandwidth of 
the nonlinear crystal is sufficiently broad, and contribution of the phase matching function is negligible.} % ignored the contribution of the phase matching amplitude of the nonlinear crystal. 
Since the signal and idler photons possess frequency correlation, detecting one photon projects the other photon~(HSP) into a mixed state. 
The frequency correlation can be removed by performing the spectral filtering or engineering the phase matching amplitude via the group velocity matching technique~(GVM)~\cite{PhysRevA.56.1534,PhysRevA.56.1627,PhysRevLett.100.133601,mosley2008conditional,PhysRevLett.105.253601} 
as shown in Fig.~\ref{fig:SPDC}(b). 
\textcolor{\nyan}{We note that, in this model, only the envelope of the pump spectrum was usually considered and its internal comb structure was neglected, since the timing 
resolution of the detectors is assumed to be enough high as was discussed in Ref.\,\cite{tapster1998}. By contrast, 
we remove the assumption to consider the influence of the timing resolution.}
%such a discrete comb structure in the pump, which is inherited by the JSA, becomes no longer negligible when the pulse interval (inverse of the frequency spacing) gets comparable to the timing jitter of the detection system, which usually occurs over the GHz repetition range.}
In this case, the JSA of the biphoton state is given by 
\begin{equation}
\Phi(\omega,\omega')\propto e^{-(\omega+\omega')^2/(2\gamma^2_p)}\sum^\infty_{n=-\infty}e^{-(\omega+\omega'-n\Delta)^2/(2\sigma^2_p)}, 
\label{eq:pump}
\end{equation}
where $\Delta$, $\gamma^2_p$ and $\sigma^2_p$ are the frequency spacing, the envelope variance of the pump comb and the variance of the each tooth, respectively.
Here, for simplicity, we \textcolor{\piyo}{considered the detuning from the center angular frequency of SPDC photons.} 
This situation is shown in Fig.~\ref{fig:SPDC}(c). 
In this regime, even after performing the GVM or the spectral filtering on the SPDC photons, the JSA still retains a frequency correlation as shown in Fig.~\ref{fig:SPDC}(d). %Namely, as one increases the pump repetition rate, the JSA gets sparser and \textcolor{\piyo}{the relevant degradation in the spectral purity of HSP} limits the clock rate of the pure HSP generation. 

\begin{figure}[t]
 \begin{center}
%\scalebox{0.27}{\includegraphics{./BellSetup.pdf}}
 \includegraphics[width=\columnwidth]{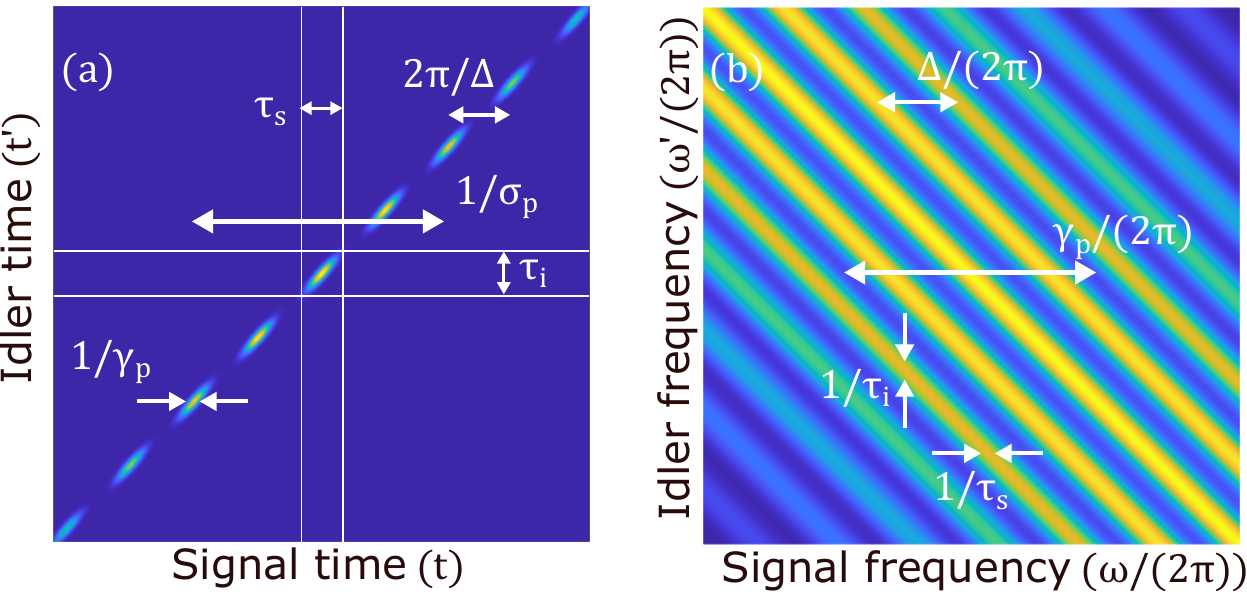}
  \caption{(a)~The JTA of the biphoton state plotted by using Eq.~(\ref{eq:timefilter}). (b)~The JSA of the biphoton state after temporal filtering plotted by using Eq.~(\ref{eq:JSA}).
  \label{fig:timefilter}} 
 \end{center}
\end{figure}

%{\it Temporal filter.-} 
\section{Temporal filtering}
We describe the role of the temporal filtering on the spectral purity.  
We consider the 2D Fourier transformation of Eq.~(\ref{eq:pump}) as  
\begin{equation}
\mathcal{F}[\Phi(\omega,\omega')]\propto e^{-(t+t')^2\sigma_p^2/2}\sum^\infty_{n=-\infty}e^{-(t+t'-2\pi n/\Delta)^2\gamma_p^2/2}\delta(t-t'), 
\label{eq:timefilter}
\end{equation}
where $\delta(t-t')$ is Dirac delta function. This joint temporal amplitude~(JTA) consists of a pulse train whose time interval is $2\pi/\Delta$, envelope variance is $1/\sigma_p^2$ and the variance of the each pulse is  $1/\gamma_p^2$. 
We extract a single pulse by introducing the temporal filter $G_{s/i}(t)$ for the signal/idler photon 
with a width of $\tau_{s/i}$ as shown in Fig.~\ref{fig:timefilter}(a). 
The JTA after the temporal filters is represented by $G_s(t)G_i(t')\mathcal{F}[\Phi(\omega,\omega')]$. 
This operation can be realized by employing narrow detection time windows using detectors with high timing resolutions. 
Performing inverse 2D Fourier transformation, we obtain the JSA after the temporal filters in angular frequency domain as 
\begin{equation}
\mathcal{F}^{-1}[G_s(t)G_i(t')]*\Phi(\omega,\omega'),  
\label{eq:JSA}
\end{equation}
where $*$ denotes the 2D convolution. The interesting property of Eq.~(\ref{eq:JSA}) is that the width of the each tooth is broadened by the temporal filters as shown in Fig.~\ref{fig:timefilter}(b). It is because the width of the each tooth is determined by the inverse of the envelope width in time domain. 
\textcolor{\piyo}{Thus the temporal filter blurs the comb structure in the pump spectrum and eliminates the correlation of the daughter photons in angular frequency domain.} 
Finally, by tailoring the envelope of the JSA using the frequency filters or the GVM technique, 
a frequency-uncorrelated pure HSP is obtained. 
The comb structure is negligible when the width of the temporal filter is smaller than the inverse of the mode spacing as $\tau_{s/i} \ll 2\pi/\Delta$. \textcolor{\piyo}{Notably, our model can quantitatively analyze the} influence of the pump repetition rate and the temporal filtering on the spectral purity of the JSA.

%In the previous HOM experiments using independent HSPs generated by pulse-pumped SPDC, the detection time windows have been set such that their widths are much larger than the coherence time of the photons. 
%Typically, the coherence time of the frequency-filtered SPDC photons are several ps, and the widths of the detection windows~($\tau_{s/i}$) have been set to a few ns. 
%When a standard mode-locked laser is used as a pump, the condition $\tau_{s/i} \ll 2\pi/\Delta$ is naturally satisfied, since 
%the typical pump repetition rate is around $f=100$\,MHz which corresponds to  $2\pi/\Delta=1/f=$10\,ns. 
%However, when one increases the pump repetition rate, narrow temporal filtering is necessary to retain the purity of the HSP. For example, when the pump repetition rate is increased to 10\,GHz, the temporal filters should be narrower than 100\,ps. 

\begin{figure*}[t]
 \begin{center}
%\scalebox{0.45}{\includegraphics{./Fig_Experiment.pdf}}
 \includegraphics[width=\columnwidth]{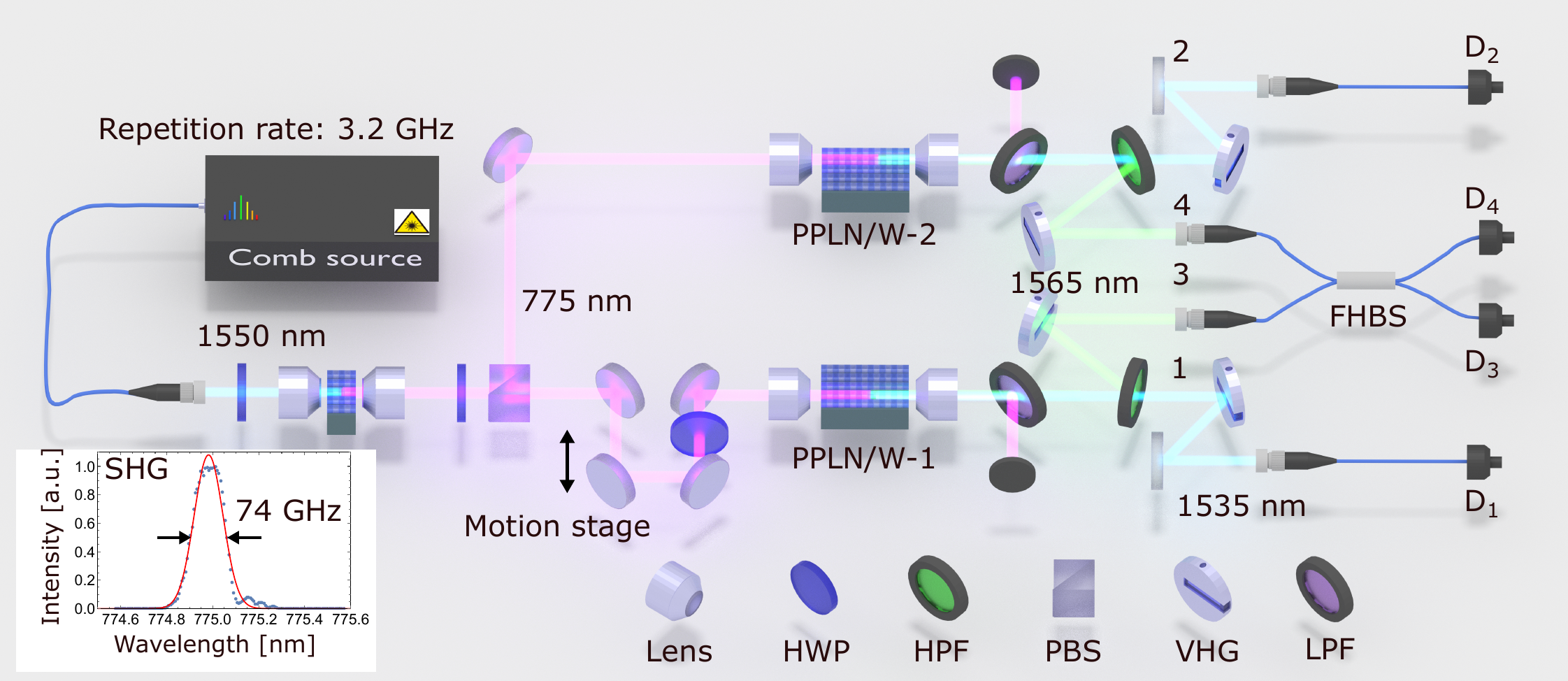}
  \caption{The experimental setup. HWP: half-waveplate, LPF: low pass filter, PBS:Polarizing beamsplitter, VHG: volume holographic grating, HPF: high pass filter, PPLN/W: PPLN waveguide. The SPDC photons are 
  directed to the SSPDs by polarization-maintaining fibers to suppress the polarization fluctuation. 
  \label{fig:Experiment}} 
 \end{center}
\end{figure*}

%{\it Experiment.-} 
\section{Experiment}
We demonstrate the HOM interference between the HSPs generated by SPDC with a \textcolor{\piyo}{3.2\,GHz-repetition-rate} mode-locked pump pulses via temporal filtering. The setup is shown in Fig.~\ref{fig:Experiment}. 
The fundamental \textcolor{\piyo}{pulse} at 1550\,nm is obtained from a  home-made \textcolor{\piyo}{frequency} comb source based on a \textcolor{\piyo}{ dual-drive electro-optical} modulator~\cite{Wakui:s}. 
The frequency of the fundamental pulses is doubled by the second harmonic generation~(SHG) using a type-0 10-mm-long periodically poled $\mathrm{LiNbO_3}$ waveguide~(PPLN/W) whose phase matching bandwidth is measured to be 174\,GHz 
full width at half maximum~(FWHM). 
\textcolor{\piyo}{The FWHM of the SHG spectrum is} $\Gamma_p=74$\,GHz as shown in the inset of Fig.~\ref{fig:Experiment}, where $2\pi\times\Gamma_{p}:=2\sqrt{\mathrm{ln}2}\gamma_p$. 
\textcolor{\piyo}{Namely, only about 23 teeth were included in the pump width.} 
The SHG pulse width is measured to be 7.4\,ps via autocorrelation.
The pump pulse is then divided into two spatial modes using a half waveplate~(HWP) and a polarizing beamsplitter~(PBS), and directed to the two SPDC sources based on type-0 34-mm-long PPLN/Ws whose phase matching bandwidths for the pump beam are measured to be 98\,GHz and 92\,GHz, respectively. 
After the strong pump beam is removed by a long pass filter~(LPF), the signal photon~(1565\,nm) and the idler photon~(1535\,nm) are divided into different spatial modes by a high pass filter~(HPF). 
We use volume holographic gratings~(VHGs) to narrow the spectral envelopes of the signal and the idler photons. 
The FWHMs of the VHGs are $\Gamma_s=32$\,GHz centered at 1565\,nm and $\Gamma_i=58$\,GHz centered at 1535\,nm, respectively. 
\begin{figure*}[t]
 \begin{center}
%\scalebox{0.17}{\includegraphics{./Fig_HOMdip.pdf}}
 \includegraphics[width=\columnwidth]{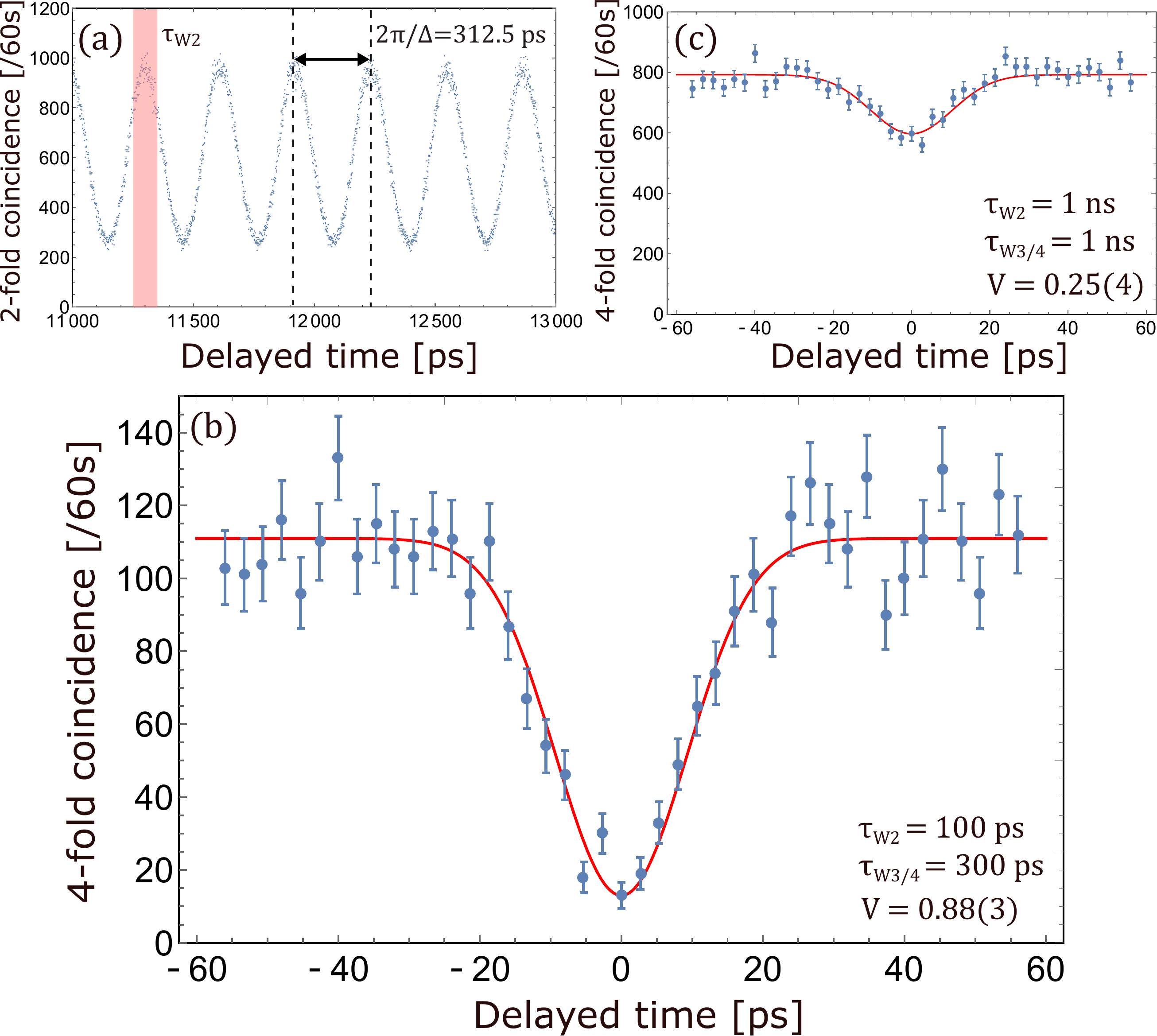}
  \caption{(a)~The accidental coincidence between $\mathrm{D}_1$ and $\mathrm{D}_2$. We employ a coincidence window with a width of $\tau_{W2}$ around 11300\,ps. 
  (b)~The HOM dip with $\tau_{1}$=148\,ps, $\tau_{W2}$=100\,ps and $\tau_{W3}=\tau_{W4}$=300\,ps. 
  (c)~The HOM dip with $\tau_{1}$=148\,ps and $\tau_{W2}=\tau_{W3}=\tau_{W4}$=1\,ns. 
  The error bars in (b) and (c) are calculated by assuming the Poissonian distribution. 
  \label{fig:HOMdip}} 
 \end{center}
\end{figure*} 
The signal photons in modes~3 and~4 are coupled to the polarization maintaining fibers~(PMFs), and mixed by a PMF-based half beamsplitter. 
All the photons are detected by superconducting \textcolor{\piyo}{nanowire} single-photon detectors~(SSPDs) \textcolor{\piyo}{\cite{Miki:17}}, 
whose timing jitters are measured to be $\tau_1=93$\,ps~($\mathrm{D}_1$), $\tau_2=148$\,ps~($\mathrm{D}_2$), $\tau_3=141$\,ps~($\mathrm{D}_3$) and $\tau_4=162$\,ps\,($\mathrm{D}_4$) FWHM, respectively. 
The electric signals from SSPDs go to a \textcolor{\piyo}{time-to-digital converter~(HydraHarp\,400, PicoQuant) whose minimum time-bin width is 1\,ps}. 
The electric signal from $\mathrm{D}_1$ is used as a start signal, and the other three electric signals are used as stop signals.  
We employ a coincidence window with a width of $\tau_{Wj}$ for each 
stop signal. Thus, the widths of the temporal filters become 
$\tau_1$ for the photon in mode~1, and the 
larger of $\tau_j$ and $\tau_{Wj}$ for the photon in mode $j=2,3,4$.

Before performing the HOM experiment, we characterize our SPDC sources. We set the average pump power coupled to PPLN/W-1 and PPLN/W-2 to be 0.32\,mW and 0.13\,mW, respectively. 
At this pump condition, the detection rates of the photon pair are $1.51\times10^5$\,cps for PPLN/W-1 and $1.35\times10^5$\,cps for PPLN/W-2, respectively. The typical Klyshko efficiency~\cite{klyshko1980use} is around 15\,$\%$ for the idler photon and 7\,$\%$ for the signal photon. \textcolor{\piyo}{The discrepancy of the efficiencies mainly comes from the difference in the VHG bandwidths.} 
To evaluate the single-photon nature of the HSPs, we measured the intensity correlation function $g^{(2)}(0)$ of the HSPs. 
$g^{(2)}(0)$ of the signal photon heralded by $\mathrm{D}_i$ for $i=$1, 2 is given by 
\begin{equation}
g^{(2)}_{i}=\frac{N(\mathrm{D}_i)N(\mathrm{D}_i\cap \mathrm{D}_3\cap D_4)}{N(\mathrm{D}_i\cap \mathrm{D}_3)N(\mathrm{D}_i\cap \mathrm{D}_4)},    
\end{equation}
where $N(\mathrm{D}_i)$ is the number of the heralding signals, and $N(\mathrm{D}_i\cap \mathrm{D}_j)$ is the coincidence count between $\mathrm{D}_i$ and $\mathrm{D}_j$. From the experimental data, $g^{(2)}_{1}$ and $g^{(2)}_{2}$ are calculated to be $2.98(9)\times10^{-2}$ and $2.71(9)\times10^{-2}$, respectively, which indicates that the amount of the higher-order photons is very small.

Finally, we perform the HOM experiment. 
To observe the HOM dip, 
we employ the coincidence window for each detection signal. For example, the accidental coincidence between $\mathrm{D}_1$ and $\mathrm{D}_2$ is shown in Fig.~\ref{fig:HOMdip}(a). 
\textcolor{\piyo}{Each peak is separated by the interval of 312.5\,ps corresponding to 3.2\,GHz pulse train and} is not perfectly separated due to the timing jitters of the SSPDs which are larger than the coherence time of SPDC photons. 
We employ a coincidence window with a width of $\tau_{W2}=100$\,ps around the corresponding peak. 
This means that we employ a 100-ps temporal filter for the idler photon in mode~2. 
For the electric signals from $\mathrm{D}_3$ and $\mathrm{D}_4$, we employ the coincidence windows with widths of $\tau_{W3}=\tau_{W4}=300$\,ps. %These relatively-large coincidence windows are necessary to observe a HOM dip, otherwise the signal peaks corresponding to the detection signal of the HSPs will protrude from the coincidence window when the relative delay gets larger. 
\textcolor{\piyo}{These relatively-large coincidence windows are 
employed to keep the 2-fold coincidence between $\mathrm{D}_1$ and $\mathrm{D}_3(\mathrm{D}_4)$ constant, otherwise the coincidence will decrease for large relative delay.}
The observed HOM dip is shown in Fig.~\ref{fig:HOMdip}(b). The visibility is $V=0.88(3)$, which is 
much higher than the classical limit of 0.5, and comparable to those observed in \textcolor{\piyo}{previous HOM experiments using conventional mode-locked lasers with repetition rates in the MHz range.} 
When we broaden the coincidence windows, the visibility is significantly degraded. We show 
the HOM dip with $\tau_{W2}=\tau_{W3}=\tau_{W4}=1$\,ns in Fig.~\ref{fig:HOMdip}(c).
While the coincidence rate increases, the visibility decreases to $V=0.25(4)$. 
\textcolor{\piyo}{In time domain, the reduced visibility is due to the fact that $\tau_{W2} > 2\pi/\Delta$ and thus pairs 
which are not coincident are wrongly considered  as ``coincident''. 
In the frequency domain, the above effect emerges as a comb structure in the JSA. }

\begin{table}[htb]
\begin{center}
  \begin{tabular}{|c|c|c|c|c|c|c|} \hline
 $\sigma_p$& $\Gamma_p$ & $\Delta/(2\pi)$ &$\tau_1$ & $\tau_\mathrm{W3}$ & $\Gamma_{s}$ & $\Gamma_{i}$  \\ \hline 
 0.5\,GHz&74\,GHz & 3.2\,GHz &148\,ps & 300\,ps &  32\,GHz  & 58\,GHz\\ \hline
   \end{tabular}
    \caption{
    \textcolor{\piyo}{Simulation parameters for Fig.~\ref{fig:JSAsim}, which are experimental values except for $\sigma_p$. Resolution (grid size) and total domain for computation were 0.1\,GHz and 300\,GHz square (larger than the plot range), respectively.}}
\label{table:JSAsim}
\end{center}
\end{table}

\begin{figure}[t]
 \begin{center}
 %\scalebox{0.5}{\includegraphics{./Fig_JSAsimulation.pdf}}
\includegraphics[width=\columnwidth]{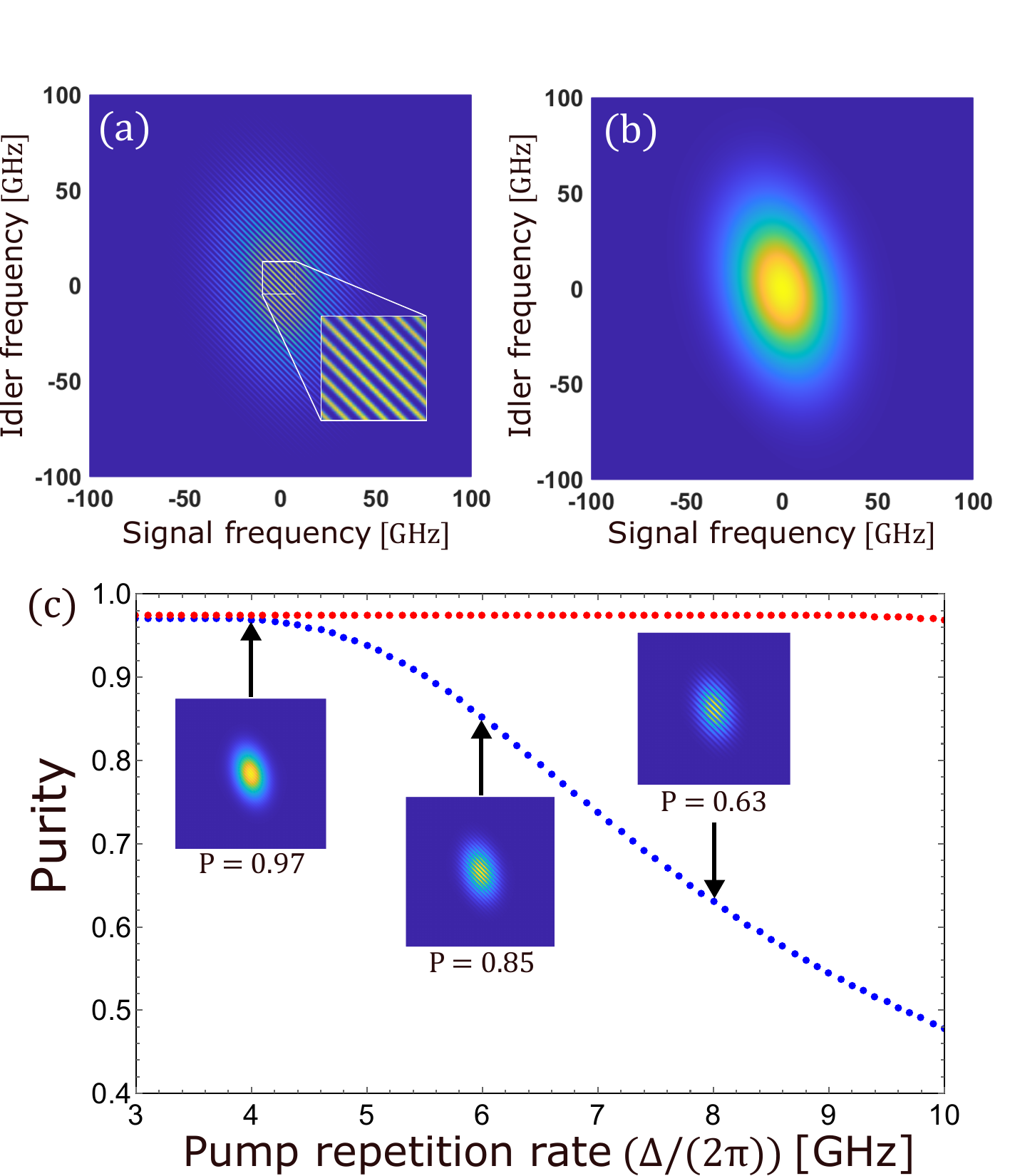}
  \caption{The simulated JSI with current experimental parameters (a)~without temporal filters and (b)~with temporal filters. (c)~The purity vs. pump repetition rate. The insets show the JSI with  $\Delta/(2\pi)=$\SI{4}{GHz}, \SI{6}{GHz} and \SI{8}{GHz}.
  \label{fig:JSAsim}} 
 \end{center}
\end{figure}

%{\it Simulation on the JSA.-}
\section{Simulation}
In this section, we calculate the spectral purity $P$ by numerical simulation using experimental parameters. 
Since the transmission functions of the spectral filters and timing jitters of detectors are experimentally obtained for the light intensity, we consider joint spectral intensity~(JSI) rather than JSA in this simulation. 
First, we consider the case without temporal filtering. 
The JSI after the filters is defined by $F_s(\omega)F_i(\omega')|\Phi(\omega,\omega')|^2$, 
where $F_{s/i}(\omega)$ is the transmission function of the spectral filters for signal/idler photon, respectively. 
We assume that $F_{s/i}(\omega)$ is a Gaussian function whose FWHM is $\Gamma_{s/i}$, respectively. 
Substituting the parameters in Table\,\ref{table:JSAsim}, we obtain the JSI as shown in Fig.~\ref{fig:JSAsim}(a). 
We clearly see the comb structure of the pump inside the JSI.
Performing the singular value decomposition on the JSI, $P$ is estimated to be $< 0.5$, which indicates that the HOM visibility cannot exceed the classical limit without temporal filtering. 
Note, $\sigma_p$ is intentionally large to emphasize each teeth in Fig.~\ref{fig:JSAsim}(a). Hence the purity can be even worse when a realistic comb width ($\sim$\,kHz) is taken into account.
Next, we consider the case with temporal filtering, where the JSI is given by $\mathcal{F}^{-1}[G_s(t)G_i(t')]*F_s(\omega)F_i(\omega')|\Phi(\omega,\omega')|^2$. 
Here, we assumed that $G_{s}(t)$ is a Gaussian function whose FWHM is $\tau_{1}$ and $G_{i}(t)$ is a rectangular function with a width of $\tau_{W3}$. 
In Fig.~\ref{fig:JSAsim}(b), the comb structure in the JSI is eliminated and the purity, which therefore became unaffected by $\sigma_p$, is calculated to be 0.97.
That is to say, the temporal filter purified the JSI.
We see that the envelope of the JSI is slightly asymmetric. The reason is that the bandwidths of the spectral filters are different between signal and idler photons. 
Finally, we scan the pump repetition rate~($\Delta/(2\pi)$) 
with all the other parameters in Table\,\ref{table:JSAsim} fixed, and plot $P$ depending on $\Delta/(2\pi)$ 
as blue dots in Fig.~\ref{fig:JSAsim}(c). 
The purity starts to decrease around $\Delta/(2\pi)=$\SI{4.5}{GHz}, and becomes lower than 0.5 for $\Delta/(2\pi)>$\SI{9.7}{GHz}. 
The red dots show the case for reference, where the timing jitters are improved to \SI{100}{ps}~($\tau_1=\tau\u{W3}$=100\,ps).
It is noteworthy that HSPs can be generated even at \SI{10}{GHz} in the spectrally pure state. In this regime, however, 
the coherence time of the HSP is comparable to the pump pulse interval, and therefore the conventional HOM dip cannot be observed. 
In such a case, the HOM visibility may be determined by the method used in cw regime~\cite{Tsujimoto2017}.

\section{Discussion} 
%{\it Discussion and conclusion.-}
We discuss the way to improve the visibility and the coincidence rate. 
We first consider the reason for the degradation of the visibility. We take the following two factors into account: (i)~the spectral purity \textcolor{\piyo}{$P$ of the HSP} and (ii)~higher-order pair creation at the SPDC process. Denoting the average photon number of the signal photons heralded by $\mathrm{D}_{1/2}$ by $s_{1/2}$, we derive a useful relation among the HOM interference visibility, $P$, $s_{1/2}$ and $g^{(2)}_{1/2}$ as
\begin{equation}
V_\mathrm{th}=P\times\frac{1}{1+\frac{\zeta g^{(2)}_1+\zeta^{-1}g^{(2)}_2}{2}}, 
\label{eq:Vtheory}
\end{equation}
where $\zeta:=s_1/s_2$. 
Here, we assumed that the JSIs are the same between the photon pairs from PPLN/W\textcolor{\piyo}{-}1 and PPLN/W\textcolor{\piyo}{-}2. 
The detailed derivation is given in Appendix~\ref{app:B}. Using the experimental parameters and results, we obtain $P=0.97$, $\zeta=1.12$, $g^{(2)}_{1}=2.98\times10^{-2}$ and $g^{(2)}_{2}=2.71\times10^{-2}$, which leads to $V_\mathrm{th}=0.94$. 
%Here, $P$ is estimated by the numerical simulation using the experimental parameters. 
The experimental value of $V=0.88(3)$ is slightly lower than this value. 
We guess the deviation from $V\u{th}$ comes from the side lobes of the VHG diffraction spectra, which is not considered in the simulation. 
We \textcolor{\piyo}{chose} the narrow VHGs such that their bandwidths are narrower than the pump bandwidth to obtain pure HSPs. 
\textcolor{\piyo}{Thus, a broader pump bandwidth may allow us to use standard band pass filters having broader bandwidths without side lobes, that should enable higher $V$.
The pump bandwidth can be broadened with a higher modulation frequency than 3.2\,GHz, while the same repetition rate is maintainable, e.g. by rate conversion using optical gating~\cite{Ishizawa2011}.}
In addition, it is necessary to replace the PPLN/Ws by those with shorter crystal lengths, otherwise the 
effective pump bandwidth is restricted by the phase matching bandwidths of the PPLN/Ws. 

In view of scalability, the coincidence rate is of importance. 
In our setup, the temporal filtering reduces the coincidence rate, since their widths are comparable to the timing jitters of the SSPDs as shown in Fig.~\ref{fig:HOMdip}(a). This effect would be negligible once photon detectors with lower timing jitters are employed. 
\textcolor{\piyo}{Recently, a single-flux-quantum coincidence circuit for SSPDs achieved} the timing jitter of 32.3\,ps~\cite{Miki2018}, \textcolor{\piyo}{presumably enabling the temporal filtering without reducing} the coincidence rate. 
\textcolor{\nyan}{In addition, the current detection rate is limited by the dead-time of SSPDs, 
since the detection efficiency of the SSPD begins to decrease around the single-count rate of $\sim2$\,MHz. 
This would be improved by employing arrayed detectors such as multi-pixel SSPD array\,\cite{Zhang2019}. 
}

\section{Conclusion} 
In conclusion, we have demonstrated the high-visibility HOM interference between two independent HSPs generated by SPDC with \textcolor{\piyo}{3.2\,GHz-repetition-rate} mode-locked pump pulses. 
%We theoretically and experimentally \textcolor{\piyo}{elucidated the mechanism by which high purity HSPs can be obtainable via temporal filtering, even when the internal structure of the sparse pump comb cannot be ignored as in the GHz range.} 
\textcolor{\nyan}{The degradation of the visibility is well explained by the theoretical model considering the pump repetition rate and the timing resolution of the
singe-photon detectors.}
The repetition rate of the pump laser is limited by the timing jitters of SSPDs. Thus, by introducing state-of-the-art photon detection system with timing jitter of several tens of ps, the high visibility HOM interference with pump repetition rate higher than 10\,GHz would be achievable. 
Combined with the techniques to efficiently generate and detect the pure HSPs, our method paves the way to a high-fidelity and high-speed photonic quantum information processing.

\appendix
\section{Purity of the HSP}
\label{app:A}
We describe how to evaluate the spectral purity of the HSP. 
The biphoton state generated by SPDC is given by 
\begin {equation}
\ket{\Psi}=\iint d\omega d\omega'\Phi(\omega,\omega')\hat{a}_s^\dagger(\omega)\hat{a}_i^\dagger(\omega')\ket{\mathrm{vac}}, 
\label{eq:1}
\end{equation}
where $\omega/\omega'$ is an angular frequency of the signal/idler photon,  $\Phi(\omega,\omega')$ is the JSA of the biphoton state, 
 and $\ket{\mathrm{vac}}$ is a vacuum state. $\hat{a}_{s/i}^\dagger(\omega)$ is the photon creation operator of the signal/idler photon whose angular frequency is $\omega$. 
We consider the case where the signal and idler photons are separated into different spatial modes, and thus the commutation relation is given by $[\hat{a}_j(\omega),\hat{a}_k^\dagger(\omega')]=\delta_{jk}\delta(\omega-\omega')$ for $j,k\in\{s,i\}$. The JSA is decomposed into  $\Phi(\omega,\omega')=\alpha(\omega,\omega')\beta(\omega+\omega')$, where $\alpha(\omega,\omega')$ and $\beta(\omega+\omega')$ are the phase matching amplitude of the nonlinear crystal and the pump spectral amplitude, respectively. 
For simplicity, we assume that the phase matching bandwidth of 
the nonlinear crystal is sufficiently broad in Sec.~\ref{sec1}, which indicates $\Phi(\omega,\omega')\simeq \beta(\omega+\omega')$.
The spectral purity of the HSP is 
determined by the factorability of the JSA, which can be tested by applying Schmidt decomposition~\cite{PhysRevLett.100.133601} on Eq.~(\ref{eq:1}) as 
\begin{equation}
    \ket{\Psi}=\sum_j\sqrt{\lambda_j}\ket{\psi_j}\ket{\phi_j}, 
\end{equation}
where $\sqrt{\lambda_j}$ is known as Schmidt coefficient which satisfies $\sum_j\lambda_j=1$, and
$\ket{\psi_j}$ and $\ket{\phi_j}$ are the orthonormal basis vectors. 
The purity of the HSP is given by $P=\sum_j\lambda^2_j$. 
If $P=1$, the HSP is pure and the spectral correlation between the signal and the idler photons is eliminated.

\section{Derivation of Eq.~(5)}
\label{app:B}
\begin{figure}[t]
 \begin{center}
%\scalebox{0.19}{\includegraphics{./Fig_Supplementary.pdf}}
 \includegraphics[width=\columnwidth]{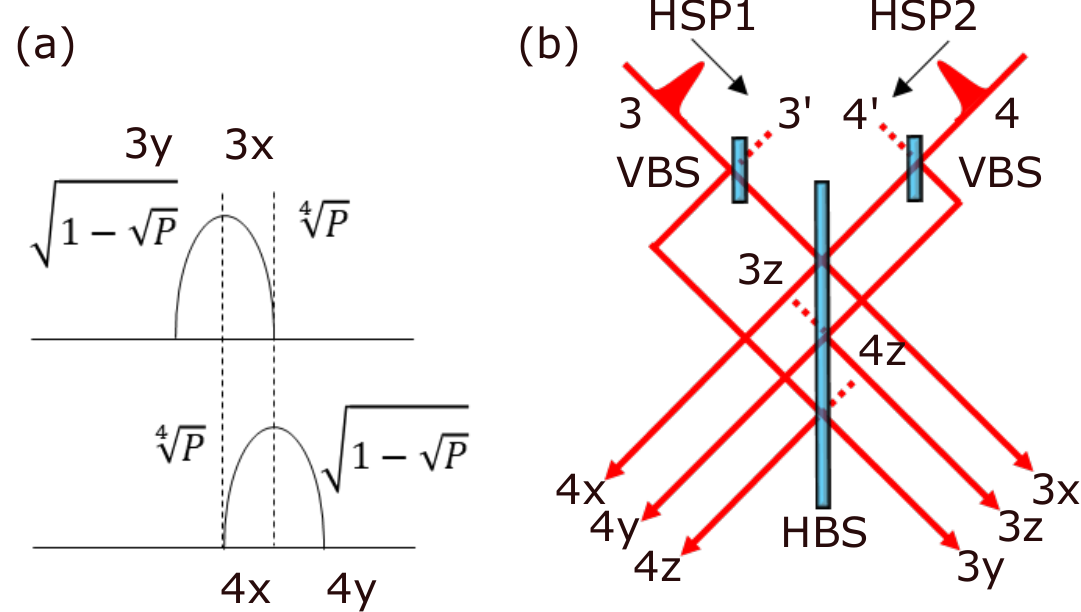}
  \caption{(a)~The mode division of the HSPs. (b)~The sketch of the HOM interference between HSP1 and HSP2. The mode of each HSP is divided by a virtual beamsplitter~(VBS). 
  \label{fig:Supplementary}} 
 \end{center}
\end{figure}

We derive the relation among the HOM visbility~($V_\mathrm{th}$), and the average photon numbers~($s_{1/2}$), the intensity correlation functions~($g^{(2)}_{1/2}$) and purity~($P$) of the HSPs. 
We assume that the JSIs of the HSP1 and HSP2 are the same, which implies that the HSPs are in single mode with probability $P$ and in the other modes where no interference occurs with probability $1-P$. This is implemented by dividing the wavepacket of each HSP into the single-mode part~($3(4)x$) with probability amplitude of $\sqrt[4]{P}$ and the multi-mode part~($3(4)y$) with probability amplitude of $\sqrt{1-\sqrt{P}}$ as shown in Fig.~\ref{fig:Supplementary}(a). The theoretical model is shown in Fig.~\ref{fig:Supplementary}(b). Each of HSP1 in modes $3$ and HSP2 in mode $4$ is divided into the two parts by the virtual beamsplitter~(VBS) whose transmittance and the reflectance are $T=\sqrt{P}$ and $R=1-\sqrt{P}$, respectively, and mixed by a HBS. 
We define the creation operators of input/output light of the HBS as $\hat{b}^{\dagger}_{ij}$/$\hat{c}^{\dagger}_{ij}$, respectively, 
where $i=3,4$ and $j=x,y,z$, and input light of the VBS as 
$\hat{a}^{\dagger}_{k}$ where $k=3,3',4,4'$.
The above operators satisfy the
commutation relations 
$[\hat{a}_{k}, \hat{a}^\dagger_{k'}]=\delta_{kk'}$, 
$[\hat{b}_{ij}, \hat{b}^\dagger_{i'j'}]=\delta_{ii'}\delta_{jj'}$ and 
$[\hat{c}_{ij}, \hat{c}^\dagger_{i'j'}]=\delta_{ii'}\delta_{jj'}$. 
Since the input states from $3'$, $4'$, $3z$ and $4z$ are vacuum states, 
$\expect{\hat{a}^\dagger_{3'}\hat{a}_{3'}}=\expect{\hat{a}^\dagger_{4'}\hat{a}_{4'}}=\expect{\hat{a}^\dagger_{3z}\hat{a}_{3z}}=\expect{\hat{a}^\dagger_{4z}\hat{a}_{4z}}=0$ holds. Assuming that the Klyshko efficiency $\eta_i$ of the system including the detection efficiency of $D_i$ for $i=3,4$ is much less than 1 
such that the detection probabilities are proportional to the photon number 
in the detected mode, the coincidence probability $C$ between $\mathrm{D}_3$ and $\mathrm{D}_4$ is given by 

\begin{equation}
C\propto\eta_3\eta_4\expect{:(\hat{n}_{3x}+\hat{n}_{3y}+\hat{n}_{3z})(\hat{n}_{4x}+\hat{n}_{4y}+\hat{n}_{4z}):}, 
\label{eq:Prob}
\end{equation}
where $\hat{n}_{ij}:=\hat{c}^\dagger_{ij}\hat{c}_{ij}$ is the number operator of the output modes. 
As is often the case in practical settings, we assume that 
the the photons in different modes have no phase correlation 
and are statistically independent. 
The HOM visibility $V\u{th}$ is calculated as follows. 
When the relative delay of the HPSs is zero, $C$ takes the minimum value of $C_0$. When the relative delay is enough large, we set $P=0$ for the VBSs, and $C$ takes the value of $C_\infty$. The HOM visibility is given by $1-C_0/C_\infty$. 
Performing the unitary transformation $\hat{U}\u{HBS}$ of the HBS follwed by 
the unitary transformation $\hat{U}\u{VBS}$ of the VBS on Eq.~(\ref{eq:Prob}), we obtain
%\begin{widetext}
\begin{equation}
C_0\propto\eta_3\eta_4\left(\expect{(\hat{a}^\dagger_{3})^2(\hat{a}_{3})^2}+\expect{(\hat{a}^\dagger_{4})^2(\hat{a}_{4})^2}+2(1-P)\expect{\hat{a}^\dagger_{3}\hat{a}^\dagger_{4}\hat{a}_{3}\hat{a}_{4}}\right)/4.     
\label{eq:Prob2}
\end{equation}
%\end{widetext}
Here, $\hat{U}\u{HBS}$ satisfies 
$\hat{U}\u{HBS}\hat{c}^\dagger_{3x}\hat{U}\u{HBS}^\dagger
=(\hat{b}^\dagger_{3x}+\hat{b}^\dagger_{4x})/\sqrt{2}$ and  
$\hat{U}\u{HBS}\hat{c}^\dagger_{4x}\hat{U}\u{HBS}^\dagger
=(\hat{b}^\dagger_{3x}-\hat{b}^\dagger_{4x})/\sqrt{2}$ for mode $x$, 
and $\hat{U}\u{HBS}\hat{c}^\dagger_{3y(z)}\hat{U}\u{HBS}^\dagger
=(\hat{b}^\dagger_{3y(z)}+\hat{b}^\dagger_{4z(y)})/\sqrt{2}$ and  
$\hat{U}\u{HBS}\hat{c}^\dagger_{4y(z)}\hat{U}\u{HBS}^\dagger
=(\hat{b}^\dagger_{3z(y)}-\hat{b}^\dagger_{4y(z)})/\sqrt{2}$ for modes
$y$ and $z$, respectively. 
$\hat{U}\u{VBS}$ satisfies 
$\hat{U}\u{VBS}\hat{b}^\dagger_{3(4)x}\hat{U}\u{VBS}^\dagger
=(\sqrt[4]{P}\hat{a}^\dagger_{3(4)}+\sqrt{1-\sqrt{P}}\hat{a}^\dagger_{3'(4')})$ and 
$\hat{U}\u{VBS}\hat{b}^\dagger_{3(4)y}\hat{U}\u{VBS}^\dagger
=(\sqrt[4]{P}\hat{a}^\dagger_{3(4)}-\sqrt{1-\sqrt{P}}\hat{a}^\dagger_{3'(4')})$, respectively. 
Using the definitions $\expect{(\hat{a}^\dagger_{3})^2(\hat{a}_{3})^2}/s_1^2=g^{(2)}_1$, 
$\expect{(\hat{a}^\dagger_{4})^2(\hat{a}_{4})^2}/s_2^2=g^{(2)}_2$ and 
$\expect{\hat{a}^\dagger_{3}\hat{a}^\dagger_{4}\hat{a}_{3}\hat{a}_{4}}=s_1s_2$, 
Eq.~(\ref{eq:Prob2}) is simplified to
\begin{equation}
C_0\propto\eta_3\eta_4\left(s_1^2g^{(2)}_2+s_2^2g^{(2)}_2+2(1-P)s_1s_2\right)/4.     
\label{eq:C0}
\end{equation}
On the other hand, $C_\infty$ is obtained by substituting $P=0$ into Eq.~(\ref{eq:C0}) as
\begin{equation}
C_{\infty}\propto\eta_3\eta_4\left(s_1^2g^{(2)}_2+s_2^2g^{(2)}_2+2s_1s_2\right)/4.     
\label{eq:Cinfty}
\end{equation}
Finally, the HOM visibility is given by 
\begin{equation}
V_\mathrm{th}=1-\frac{C_0}{C_\infty}=P\times\frac{1}{1+\frac{\zeta g^{(2)}_1+\zeta^{-1}g^{(2)}_2}{2}}, 
\end{equation}
where $\zeta:=s_1/s_2$.

\section*{Funding}
Core Research for Evolutional Science and Technology~(JPMJCR1772); Japan Society for the
Promotion of Science~(JP18K13487, JP20K14393). 

\section*{Acknowledgments.}
Y.T. thanks Rikizo Ikuta for helpful discussions. 

\section*{Disclosures.}
The authors declare no conflicts of interest. 

\section*{Data availability.}
Data underlying the results presented in this paper are not publicly available at this time but may
be obtained from the authors upon reasonable request. 

%\bibliography{ref}

% \begin{thebibliography}{10}
\bibliographystyle{osajnl}

\end{document}